\documentclass[
    ,final            
  ]
  {aipproc}

\usepackage{amsmath}
\usepackage{color}
\usepackage{xfrac}
\layoutstyle{8x11single}

\begin{document}

\title{Dark matter on the lattice}


\author{Randy Lewis}{
  address={Department of Physics and Astronomy, York University, Toronto, Ontario, Canada M3J 1P3}
}

\begin{abstract}
Several collaborations have recently performed lattice calculations aimed
specifically at dark matter, including work with SU(2), SU(3), SU(4) and
SO(4) gauge theories to represent the dark sector.
Highlights of these studies are presented here, after
a reminder of how lattice calculations in QCD itself are helping with
the hunt for dark matter.
\end{abstract}

\maketitle

During the past 3 years, there have been exploratory lattice studies of dark matter with
an SU(2) gauge theory \cite{Lewis:2011zb,Hietanen:2014xca,Hietanen:2013fya,Detmold:2014qqa,Detmold:2014kba},
an SU(3) gauge theory \cite{Appelquist:2013ms},
an SU(4) gauge theory \cite{Appelquist:2014jch} and
an SO(4) gauge theory \cite{Hietanen:2012sz}.
Before presenting these, let's recall how lattice QCD has determined a
matrix element that is valuable to a broad range of dark matter models.

\section{Lattice QCD studies for dark matter}

For a valuable review of this topic, including many details and
references to the extensive original literature, see
Ref.~\cite{Junnarkar:2013ac} by Junnarkar and Walker-Loud.
The discussion presented below is merely an appetizer.

If dark matter is a weakly-interacting massive particle (WIMP), then
experimental observation requires knowledge of how the WIMP interacts with
a detector.  Therefore the basic WIMP-nucleon interaction is of fundamental
interest.  Although the microscopic theory of WIMPs is unknown,
the low-energy limit of a generic spin-independent interaction is scalar
and that scalar could couple to any flavor of quark
in the nucleon (not only valence flavors).  Up/down quark contributions can be
extracted from pion-nucleon scattering experiments,
charm/bottom/top quark contributions are
amenable to perturbation theory, and the strange quark contribution has
been obtained from lattice QCD calculations.

There are no valence strange quarks in a nucleon and
it is notoriously difficult to calculate with lattice QCD in a situation
like this one, where the scalar probe is coupling to a virtual quark rather
than a valence quark.
Nevertheless, many collaborations have accepted the challenge \cite{JWplot}.
The matrix element studied in lattice QCD is
\begin{equation}\label{strange}
f_s = \frac{\langle N|m_s\bar{s}s|N\rangle}{m_N}
    = \frac{m_s}{m_N}\frac{\partial m_N}{\partial m_s}
\end{equation}
where the first equality defines the dimensionless ratio $f_s$ and the second
equality is the Feynman-Hellman theorem.

\vspace{5mm}
\begin{center}
{\em Click the box to see Fig.~8 from Ref.~\cite{Junnarkar:2013ac}, showing 17
determinations of $f_s$ from the lattice QCD community.}

{\color{blue}\framebox{\framebox{\url{http://inspirehep.net/record/1209574/files/figures_fs_compare.png}}}}
\end{center}
\vspace{5mm}

The lattice QCD implementations differ from one another in several ways, and
this provides confidence in understanding systematic effects.
Some calculations use the Feynman-Hellman theorem while others calculate the
matrix element directly.  Different lattice spacings, lattice volumes and
lattice actions are used.  Different methods of reaching the physical
up/down masses are employed.  The color coding in the figure reflects the
confidence that Junnarkar and Walker-Loud recommend for individual
determinations of $f_s$ but, regardless of these colors, the lattice results
display a clear indication that $f_s$ is significantly smaller than 0.1.
The authors of Ref.~\cite{Junnarkar:2013ac} arrive at a lattice average of
\begin{equation}
f_s = 0.043\pm0.011
\end{equation}
which is valuable input for phenomenological studies of dark matter.

\section{Putting dark matter directly on the lattice}

Dark matter is physics beyond the standard model.
If that physics comes with a new strong interaction, then lattice field theory
could be a valuable calculational tool.
Each section below discusses lattice studies performed for a particular
candidate gauge theory in the dark sector.
The discussion here will be very brief, and readers should consult
Refs.~\cite{Lewis:2011zb,Hietanen:2014xca,Hietanen:2013fya,Detmold:2014qqa,Detmold:2014kba,Appelquist:2013ms,Appelquist:2014jch,Hietanen:2012sz} 
for important details and for references to the large
body of phenomenological literature on which many of the underlying ideas
are based.

\subsection{SU(2) gauge theory}

As a smaller cousin of QCD, SU(2) is a natural starting
point for lattice explorations of possible gauge theories in the dark sector.
A convenient matter content is provided by
having just 2 Dirac fermions, both in the fundamental representation.
As verified by the explicit lattice calculations in Ref.~\cite{Lewis:2011zb},
this theory has a global SU(4) symmetry that is dynamically broken to Sp(4),
and the corresponding 5 Goldstone bosons are observed in the spectrum.
They couple to the following operators corresponding to 3 dark pions, 1
dark nucleon and 1 dark anti-nucleon, which are all spinless in SU(2):
\begin{eqnarray}
\Pi^+ &\Leftrightarrow& \bar{U}\gamma_5D \,,\\
\Pi^- &\Leftrightarrow& \bar{D}\gamma_5U \,,\\
\Pi^0 &\Leftrightarrow& (\bar{U}\gamma_5U-\bar{D}\gamma_5D)/\sqrt{2} \,,\\
\Pi_{UD} &\Leftrightarrow& U^T(-i\sigma^2C)\gamma_5D \,,\\
\Pi_{\overline{UD}} &\Leftrightarrow& \bar{U}(-i\sigma^2C)\gamma_5\bar{D}^T \,,
\end{eqnarray}
where $C$ is the charge conjugation matrix and the Pauli structure
$(-i\sigma^2)$ acts on color indices.
If the $U$ and $D$ fermions have equal non-zero masses,
then all 5 (now pseudo) Goldstone bosons acquire a common mass.

\vspace{5mm}
\begin{center}
{\em Click the box to see Fig.~4 from Ref.~\cite{Lewis:2011zb}, showing the
Goldstone boson mass squared versus fermion mass.}

{\color{blue}\framebox{\framebox{\url{http://inspirehep.net/record/927706/files/GBmass.png}}}}
\end{center}
\vspace{5mm}

The left panel of the figure confirms the expected $m_\Pi^2\propto m_q$ behavior
for Goldstone bosons.  The right panel confirms the expected deviations due to
finite volume and due to large $m_q$.

Where would these 5 Goldstone bosons appear in nature?
Ref.~\cite{Lewis:2011zb} mentions the option
that the 3 dark pions could be eaten by the $W^\pm$ and $Z$ bosons while the
dark nucleon is a dark matter candidate and the dark anti-nucleon
has been annihilated away through baryon asymmetry (like the
QCD anti-nucleon).
This option would require the 125 GeV boson to emerge from the SU(2) theory
as a $0^{++}$ hadron, and corresponds to technicolor-style electroweak
couplings.

Ref.~\cite{Hietanen:2014xca} takes a more general view, allowing for an
arbitrary rotation angle $\theta$ somewhere between the technicolor limit
($\theta=\pi/2$) and a composite Higgs limit ($\theta=0$).
In the latter limit, the dark nucleon/anti-nucleon
degrees of freedom realign into a composite Higgs and a dark matter candidate.

Beyond the pseudo-Goldstone bosons, this SU(2) theory has vector and
axial-vector mesons awaiting experimental discovery.
Ref.~\cite{Hietanen:2014xca} computes the vector mass to be
\begin{equation}
m_\rho = \frac{2.5\pm0.5 {\rm ~TeV}}{\sin\theta}
\end{equation}
based on data in the following figure:

\vspace{5mm}
\begin{center}
{\em Click the box to see Fig.~5 from Ref.~\cite{Hietanen:2014xca}, showing the
vector and axial-vector meson masses versus fermion mass.}

{\color{blue}\framebox{\framebox{\url{http://inspirehep.net/record/1289883/files/spectrum_2.png}}}}
\end{center}
\vspace{5mm}

Ref.~\cite{Hietanen:2013fya} calculates the cross section for scattering
the dark matter candidate from a proton.
Contributions from photon exchange and from composite Higgs exchange are both
included but, as we'll see below, the photon
dominates when approaching current experimental bounds.

A direct lattice calculation of the photon coupling to each of the 5
Goldstone bosons is given by
\begin{eqnarray}
C^{(3)}_{UD}(t_i,t,t_f,\vec p_i,\vec p_f) &=& T^U - T^D \label{dmff} \,,\\
C^{(3)}_{\overline{UD}}(t_i,t,t_f,\vec p_i,\vec p_f) &=& -T^U + T^D \,,\\
C^{(3)}_{\Pi^+}(t_i,t,t_f,\vec p_i,\vec p_f) &=& T^U + T^D \,,\\
C^{(3)}_{\Pi^-}(t_i,t,t_f,\vec p_i,\vec p_f) &=& -T^U - T^D \,,\\
C^{(3)}_{\Pi^0}(t_i,t,t_f,\vec p_i,\vec p_f) &=& 0 \,,
\end{eqnarray}
where
\begin{equation}
T^X = \sum_{\vec x_i,\vec x,\vec x_f}e^{-i(\vec x_f-\vec x)\cdot\vec p_f}
      e^{-i(\vec x-\vec x_i)\cdot\vec p_i}
      \left<0\left|\Pi(t_f,\vec x_f)V^X_\mu(t,\vec x)\Pi^\dagger(t_i,\vec x_i)
      \right|0\right> \,.
\end{equation}
Eq.~(\ref{dmff}) vanishes if $m_U=m_D$, so the dark matter candidate has
no electromagnetic form factor at all in that limit.
Unfortunately, a lattice calculation with
$m_U\neq m_D$ is very costly due to the presence of disconnected quark
loop effects.  Ref.~\cite{Hietanen:2013fya} avoids this dilemma by using
lattice simulations at $m_U=m_D$ together with vector meson dominance.
A priori it is not known whether vector meson dominance is reliable in an
SU(2) gauge theory, so the lattice calculations in Ref.~\cite{Hietanen:2013fya}
were used to verify its applicability.

\vspace{5mm}
\begin{center}
{\em Click the box to see Fig.~9 from Ref.~\cite{Hietanen:2013fya}, showing the
applicability of vector meson dominance for these lattice data.}

{\color{blue}\framebox{\framebox{\url{http://inspirehep.net/record/1249877/files/ff3.png}}}}
\end{center}
\vspace{5mm}

The photon couples to the dark matter candidate through the charge radius,
which in this case is
\begin{equation}
\mathcal{L}_B=ie\frac{d_B}{m_\rho^2}
   \phi^*\overleftrightarrow{\partial_\mu} \phi \, \partial_{\nu}F^{\mu\nu} \,,
~~~~~{\rm where}~~ d_B=\frac{m_{\rho_U}-m_{\rho_D}}{m_\rho} \,.
\end{equation}
Notice that the charge radius vanishes when $m_U=m_D$, as expected.
The cross section obtained from photon exchange and composite Higgs exchange
with $|d_B|=1$ and $|d_B|=0.1$ (dot-dashed in the figures) is essentially
proportional to $|d_B|^2$, indicating that composite Higgs exchange is
negligible.  The cross section can be large enough to be constrained by upcoming
experimental searches.

\vspace{5mm}
\begin{center}
{\em Click the boxes to see Fig.~10 from Ref.~\cite{Hietanen:2013fya}, showing the
dark matter-proton cross section and experimental bounds.}

{\color{blue}\framebox{\framebox{\url{http://inspirehep.net/record/1249877/files/pNGB_DM2.png}}}}

{\color{blue}\framebox{\framebox{\url{http://inspirehep.net/record/1249877/files/pNGB_DM1.png}}}}
\end{center}
\vspace{5mm}

Ref.~\cite{Detmold:2014kba} calculates the scalar coupling to some of the
dark hadrons in this SU(2) theory.  It is the analogue of Eq.~(\ref{strange})
but now in the dark sector, i.e.
\begin{equation}
f_q^{(H)} = \frac{\langle H|m_q\bar{q}q|H\rangle}{m_H}
    = \frac{m_q}{m_H}\frac{\partial m_H}{\partial m_q}
\end{equation}
where $H$ denotes a dark hadron of interest.
Several calculations are done for a modest range of valence fermion masses
surrounding the sea quark mass, and the Feynman-Hellman theorem is used to arrive
at precise numerical results.
The authors of Ref.~\cite{Detmold:2014kba} expect $O(30\%)$ systematic errors
due to partial quenching.

\vspace{5mm}
\begin{center}
{\em Click the box to see Fig.~26 from Ref.~\cite{Detmold:2014kba}, showing the
determination of $f_{u+d}^{H}$ for 2 hadron multiplets.}

{\color{blue}\framebox{\framebox{\url{http://inspirehep.net/record/1300660/files/figures_fHPlot.png}}}}
\end{center}
\vspace{5mm}

Refs.~\cite{Detmold:2014qqa,Detmold:2014kba} point out that, like QCD, the SU(2) theory
could also have a nuclear physics spectrum with several nuclei contributing
to dark matter.  Lattice calculations are used to test several options,
learning whether they form nuclear bound states or are merely scattering
states.  The basic idea is to study the dependence of energy on lattice
volume $L^3$,
\begin{eqnarray}
\Delta E(L) \propto \frac{1}{L^3}+\ldots \hspace{20.4mm}
             &\Rightarrow& {\rm scattering~states} \\
\Delta E(L) =-\frac{\gamma^2}{2\mu}\left(1+\frac{12\hat C}{\gamma L}
             e^{-\gamma L}\right)  &\Rightarrow& {\rm bound~states}
\end{eqnarray}
but please consult Ref.~\cite{Detmold:2014kba} directly for a thorough
discussion.  Detmold, McCullough and Pochinsky find evidence for bound
states with $J^P=1^+$ composed of $N\Delta$ and $NN\Delta$ and perhaps also
$NNN\Delta$.  Recall that all hadrons are bosons in a 2-color theory;
$N$ and $\Delta$ denote spin 0 and spin 1 states respectively.
This study chooses fermion masses that lead to $m_\rho/2<m_\pi<m_\rho$ in
the dark sector rather than working near the chiral limit.
As in Refs.~\cite{Lewis:2011zb,Hietanen:2014xca,Hietanen:2013fya}, the
overall scale is set by $f_\pi=246$ GeV.

\vspace{5mm}
\begin{center}
{\em Click the boxes to see Fig.~23 from Ref.~\cite{Detmold:2014kba}, showing the
determination of $\gamma$ for $N\Delta$, $NN\Delta$ and $NNN\Delta$.}

{\color{blue}\framebox{\framebox{\url{http://inspirehep.net/record/1300660/files/figures_TeV2DgammaFit-1NDelta.png}}}}

{\color{blue}\framebox{\framebox{\url{http://inspirehep.net/record/1300660/files/figures_TeV2DgammaFit-2NDelta.png}}}}

{\color{blue}\framebox{\framebox{\url{http://inspirehep.net/record/1300660/files/figures_TeV2DgammaFit-3NDelta.png}}}}
\end{center}

\subsection{SU(3) gauge theory}

The lattice community has a lot of experience with SU(3) gauge theory for QCD.
Ref.~\cite{Appelquist:2013ms} considers the possibility of a new SU(3)
gauge theory in the dark sector having either 2 or 6 flavors of Dirac
fermions in the fundamental representation.  All dark fermions are taken to
be singlets under the standard model weak interaction and to come in
pairs with electric charges $Q_U=+2/3$ and $Q_D=-1/3$.
In contrast to the SU(2) case discussed above, this new SU(3) sector is not
given any role in electroweak symmetry breaking.
The $N_f^2-1$ Goldstone bosons are assumed unstable, leaving the lightest
baryon (i.e.\ the dark neutron) as the dark matter candidate.

A lattice calculation of the electromagnetic form factor is used to
obtain the event rate that would be observed in the XENON100 experiment,
as a function of the dark matter particle's mass.
Disconnected fermion loops,
presumably a small contribution, are omitted because of their excessive expense.
The authors of Ref.~\cite{Appelquist:2013ms} observe that the mean square
charge radius, $\langle r_E^2\rangle$, is an order of magnitude larger than
its QCD counterpart due to their use of a quite large mass for the
dark fermion.

\vspace{5mm}
\begin{center}
{\em Click the box to see Fig.~6 from Ref.~\cite{Appelquist:2013ms}, comparing the
XENON100 bound to a specific SU(3) dark matter model.}

\begin{small}
{\color{blue}\framebox{\framebox{\url{http://inspirehep.net/record/1209894/files/rate_XE100_6p6-43p3_12075988_full_rEonly.png}}}}
\end{small}
\end{center}
\vspace{5mm}

In the figure, solid curves are the computed event rate, while dashed
curves show the contribution from just the charge radius term.
Notice that the cross section has only a tiny sensitivity to
whether $N_f=2$ or 6.  To respect the bound from XENON100, the
dark matter particle must have a mass of more than 10 TeV.

\subsection{SU(4) gauge theory}

Ref.~\cite{Appelquist:2014jch} reports on a quenched exploration of an SU(4)
model for dark matter.
Like the SU(2) case, all hadrons are bosons.
Unlike the SU(2) case, all hadrons do not have just 2 valence fermions;
mesons still have 2 but baryons now have 4.

A first step is to compare the mass spectrum of SU(4) [with solid error bars]
to the more familiar SU(3) [with dashed errors bars].  Two options for the
fermion mass are displayed, and the horizontal axis shows that the mass ratio of
pseudoscalar to vector mesons is in the 70\%-80\% range.
The authors note that the pseudoscalar mass must be larger than 100 GeV to
satisfy LEP bounds.

The lattice results show that meson masses are largely independent of
whether $N_c=3$ or 4, as expected.
The lattice results also show that baryon masses [$J=0,1,2$ for SU(4) and
$J=\sfrac{1}{2},\sfrac{3}{2}$ for SU(3)] are roughly proportional to $N_c$, as expected.

\vspace{5mm}
\begin{center}
{\em Click the box to see Fig.~3 from Ref.~\cite{Appelquist:2014jch}, comparing the
hadron masses in SU(3) and SU(4) gauge theories.}

{\color{blue}\framebox{\framebox{\url{http://inspirehep.net/record/1282609/files/3cv4c_no_rotor.png}}}}
\end{center}
\vspace{5mm}

Ref.~\cite{Appelquist:2014jch} calculates the dark matter-nucleon scattering
cross section as a function of the coupling of the dark baryon to the Higgs,
which is named $\alpha$,
\begin{equation}
\alpha = \frac{v}{m_f}\frac{\partial m_f(h)}{\partial h}\bigg|_{h=v} \,,
~~~~~{\rm where}~~ v=246 {\rm ~GeV} \,.
\end{equation}

The following figure shows that, for one particular value of
$m_{\rm PS}/m_{\rm V}$, the coupling $\alpha$ must be quite small to be
compatible
with the LUX experimental bound.  LEP constraints require $m_B>300$ GeV.
Effects of lattice spacing, lattice volume, and a broader range of
$m_{\rm PS}/m_{\rm V}$ were also studied; please
consult Ref.~\cite{Appelquist:2014jch} for more information.

\vspace{5mm}
\begin{center}
{\em Click the box to see Fig.~5(b) from Ref.~\cite{Appelquist:2014jch}, showing
a dark matter-nucleon cross section in an SU(4) gauge theory.}

{\color{blue}\framebox{\framebox{\url{http://inspirehep.net/record/1282609/files/LUX-exclusion-k015625-label.png}}}}
\end{center}

\subsection{SO(4) gauge theory}

The first lattice study applying a non-SU(N) theory to dark matter is
Ref.~\cite{Hietanen:2012sz}, authored by Hietanen, Pica, Sannino and
S\o ndergaard.  They use 2 (Wilson) Dirac fermions in the vector
representation and begin by exploring the phase diagram through varying the 2
parameters: bare inverse gauge coupling $\beta$ and bare fermion mass $m_0$.
A physical region is found where $\beta$ and $m_0$ are both sufficiently
large.

\vspace{5mm}
\begin{center}
{\em Click the box to see Fig.~2 from Ref.~\cite{Hietanen:2012sz}, showing
the phases of a lattice SO(4) gauge theory.}

{\color{blue}\framebox{\framebox{\url{http://inspirehep.net/record/1203463/files/LatticePhases.png}}}}
\end{center}
\vspace{5mm}

The Polyakov loop is a gauge-invariant path in one spatial direction, closing
back on itself due to periodic boundary conditions.  Since the lattice
action treats every spatial direction equally, Polyakov loops are
expected to be statistically equivalent in each spatial direction.
Ref.~\cite{Hietanen:2012sz} observed an interesting counterexample in
small lattice volumes, interpreted as the emergence of two distinct phases
during the simulation.  This phenomenon was not observed for larger
volumes.

\vspace{5mm}
\begin{center}
{\em Click the box to see Fig.~6 from Ref.~\cite{Hietanen:2012sz}, showing
spatial Polyakov loops for a $12^3\times64$ lattice.}

{\color{blue}\framebox{\framebox{\url{http://inspirehep.net/record/1203463/files/L12T64beta7m03_pit.png}}}}
\end{center}
\vspace{5mm}

This theory has a global SU(4) symmetry that breaks to a global SO(4),
producing 9 Goldstone bosons.  The Goldstone boson having isospin zero is the dark
matter candidate.  The pseudoscalar, vector and axial-vector meson masses
have been calculated as a function of quark mass, and the Goldstone nature
of the pseudoscalars has been observed.

\vspace{5mm}
\begin{center}
{\em Click the box to see Fig.~8 from Ref.~\cite{Hietanen:2012sz}, showing
meson masses in an SO(4) gauge theory.}

{\color{blue}\framebox{\framebox{\url{http://inspirehep.net/record/1203463/files/L24T64b7mesons.png}}}}
\end{center}

\vspace{5mm}

\begin{center}
{\em Click the box to see Fig.~9 from Ref.~\cite{Hietanen:2012sz}, showing
the vector to pseudoscalar mass ratio for an SO(4) gauge theory.}

{\color{blue}\framebox{\framebox{\url{http://inspirehep.net/record/1203463/files/L24T64b7PS_V_ratio.png}}}}
\end{center}

\section{Summary}

Studying dark matter models with lattice field theory is a recent effort
comprising Refs.~\cite{Lewis:2011zb,Hietanen:2014xca,Hietanen:2013fya,Detmold:2014qqa,Detmold:2014kba,Appelquist:2013ms,Appelquist:2014jch,Hietanen:2012sz}
and there is a lot more that can be explored.  The most-studied model so
far is SU(2) gauge theory with 2 Dirac fermions in the fundamental
representation \cite{Lewis:2011zb,Hietanen:2014xca,Hietanen:2013fya,Detmold:2014qqa,Detmold:2014kba} that connects dark matter to electroweak symmetry breaking.

The lattice QCD determination of $\langle N|m_s\bar{s}s|N\rangle$ is
mature \cite{Junnarkar:2013ac,JWplot},
with independent studies by many collaborations and detailed accounting of
systematic uncertainties.

\begin{theacknowledgments}
Thanks to the organizers of {\em XIth Quark Confinement and the Hadron Spectrum}
for the opportunity to participate.  Thanks to Ari Hietanen, Claudio Pica and
Francesco Sannino for valuable collaboration into dark matter on the lattice.
This work was supported in part by the Natural Sciences and Engineering Research Council (NSERC) of Canada.
\end{theacknowledgments}

\end{document}